\documentclass[article,twocolumn,superscriptaddress,citeautoscript,reprint]{revtex4-2}

\usepackage{blindtext}
\usepackage{graphicx}
\usepackage{bm}
\usepackage{epsfig}
\usepackage{amssymb}
\usepackage{amsfonts}
\usepackage{braket}
\usepackage{color}
\usepackage{epstopdf}
\usepackage{hyperref}
\usepackage{float}
\restylefloat{table}
\usepackage{bibentry}
\usepackage{color}
\usepackage{multirow}
\usepackage[caption=false]{subfig}
\usepackage[normalem]{ulem}

\usepackage{amsfonts}
\usepackage{amsthm}
\usepackage{graphicx}
\usepackage{multirow}
\usepackage{color}
\usepackage{bbold}
\usepackage{bm}
\usepackage{times}
\usepackage{amsmath,bm,amsfonts}
\usepackage{dcolumn}
\usepackage{graphicx}
\usepackage{latexsym}

\newcommand{\bpm}{\begin{pmatrix}}
	\newcommand{\epm}{\end{pmatrix}}
\newcommand{\ba}{\begin{eqnarray}}
	\newcommand{\ea}{\end{eqnarray}}
\newcommand{\bd}{\begin{displaymath}}

	\graphicspath{{figures/}}% Put all figures in this directory.

\usepackage{pdfpages} % include pdfs
\usepackage{pgffor} % for loops

\makeatletter
\AtBeginDocument{\let\LS@rot\@undefined}
\makeatother

\def\supplementfilename{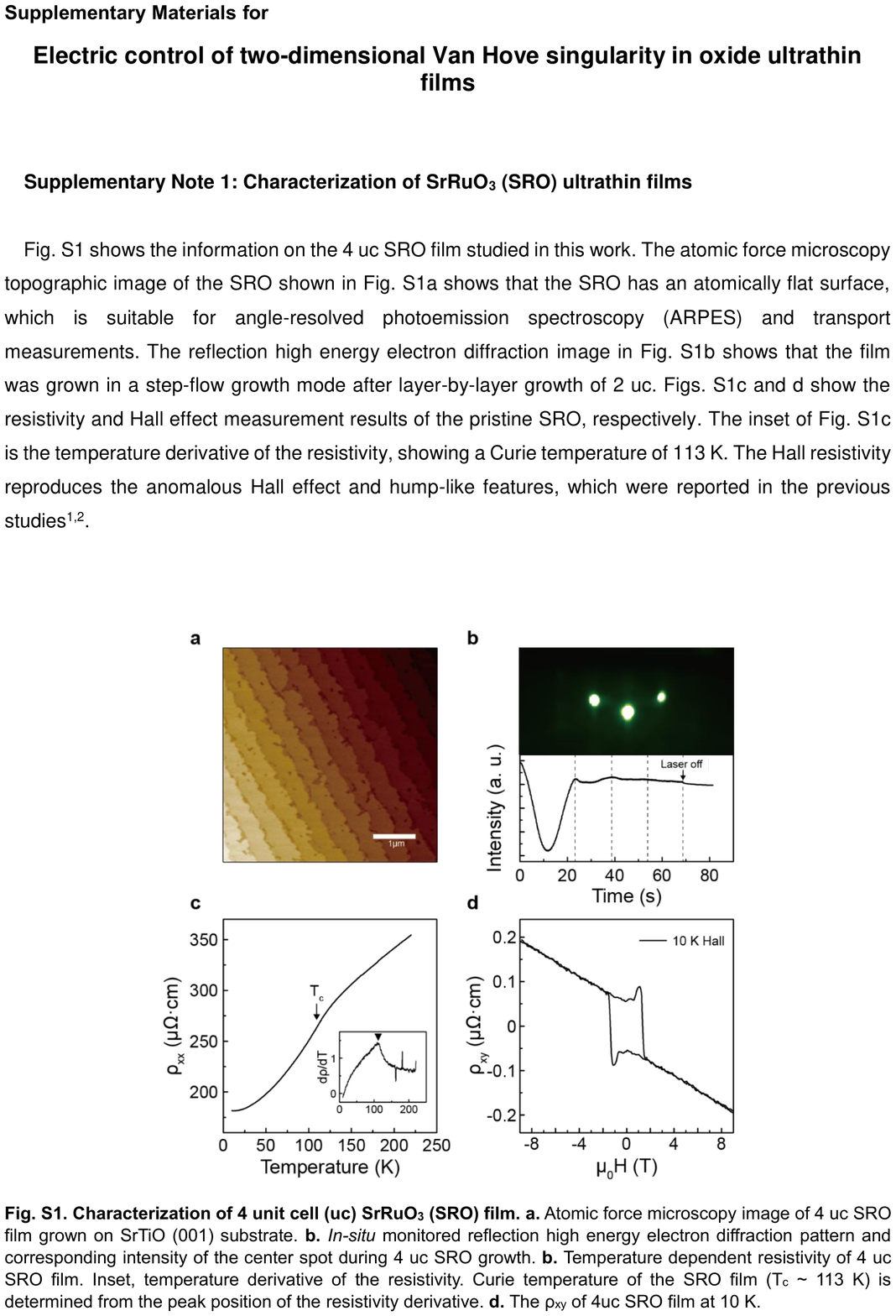}

\pdfximage{\supplementfilename}
\def\numbersupplementpages{\the\pdflastximagepages}

\newif\ifarXiv
\arXivtrue 
\usepackage{lipsum}
\usepackage{fancyhdr}
\begin{document}
	
\title{Electric control of two-dimensional Van Hove singularity in oxide ultra-thin films}

\author{Donghan Kim}
\thanks {These authors contributed equally to this work.}
\affiliation{Center for Correlated Electron Systems, Institute for Basic Science, Seoul 08826, Korea}
\affiliation{Department of Physics and Astronomy, Seoul National University, Seoul 08826, Korea}

\author{Younsik Kim}
\thanks {These authors contributed equally to this work.}
\affiliation{Center for Correlated Electron Systems, Institute for Basic Science, Seoul 08826, Korea}
\affiliation{Department of Physics and Astronomy, Seoul National University, Seoul 08826, Korea}

\author{Byungmin Sohn}
\email[Electronic address:$~~$]{sbm1000@snu.ac.kr}
\affiliation{Center for Correlated Electron Systems, Institute for Basic Science, Seoul 08826, Korea}
\affiliation{Department of Physics and Astronomy, Seoul National University, Seoul 08826, Korea}

\author{Minsoo Kim}
\affiliation{Center for Correlated Electron Systems, Institute for Basic Science, Seoul 08826, Korea}
\affiliation{Department of Physics and Astronomy, Seoul National University, Seoul 08826, Korea}

\author{Bongju Kim}
\affiliation{Center for Correlated Electron Systems, Institute for Basic Science, Seoul 08826, Korea}
\affiliation{Department of Physics and Astronomy, Seoul National University, Seoul 08826, Korea}

\author{Tae Won Noh}
\affiliation{Center for Correlated Electron Systems, Institute for Basic Science, Seoul 08826, Korea}
\affiliation{Department of Physics and Astronomy, Seoul National University, Seoul 08826, Korea}

\author{Changyoung Kim}
\email[Electronic address:$~~$]{changyoung@snu.ac.kr}
\affiliation{Center for Correlated Electron Systems, Institute for Basic Science, Seoul 08826, Korea}
\affiliation{Department of Physics and Astronomy, Seoul National University, Seoul 08826, Korea}

\date{\today}

\begin{abstract}
Divergent density of states (DOS) can induce extraordinary phenomena such as significant enhancement of superconductivity and unexpected phase transitions. Moreover, van Hove singularities (VHSs) are known to lead to divergent DOS in two-dimensional (2D) systems. Despite the recent interest in VHSs, only a few controllable cases have been reported to date. In this work, we investigate the electronic band structures of a 2D VHS with angle-resolved photoemission spectroscopy and control transport properties by utilizing an atomically ultra-thin SrRuO$_3$ film. By applying electric fields with alkali metal deposition and ionic-liquid gating methods, we precisely control the 2D VHS, and the sign of the charge carrier. Use of a tunable 2D VHS in an atomically flat oxide film could serve as a new strategy to realize infinite DOS near the Fermi level, thereby allowing efficient tuning of electric properties. 
\end{abstract}
%\pacs{}% insert suggested PACS numbers in braces on next line
\maketitle

\section*{Introduction}
Enhancing the density of states (DOS) is an important way to tune the physical properties of materials. Enhancement of the DOS near the Fermi level ($E_{\rm F}$) often leads to unexpected emergent phenomena, such as enhanced superconductivity~\cite{hicks2014strong,steppke2017strong}, density waves~\cite{rice1975new,tan2021charge}, and magnetism~\cite{tserkovnyak2005nonlocal,goodenough1967narrow}. This was exploited by a recent approach to realize flat band systems in Kagome lattices~\cite{tan2021charge,ye2018massive,kang2020topological,cho2021emergence,kang2022twofold,kang2020dirac} and twisted bilyaer graphene~\cite{lisi2021observation,cao2018unconventional}. Another promising approach is placing saddle point van Hove singularities (VHSs) near $E_{\rm F}$. Since VHSs have locally maximized DOS, novel properties as well as their control may be realized with VHSs~\cite{van1953occurrence}. Several studies have been reported on controlling VHSs and resultant change in physical properties such as superconductivity~\cite{hicks2014strong,steppke2017strong}, nontrivial Hall effects~\cite{hatsugai2006topological,gobel2017unconventional}, and metal-insulator transition~\cite{lee2021hund,bonvca2009van} in various materials.

A desirable way to control VHSs is changing the chemical potential of materials through an externally applied electric field, which induces charge carriers near the surface of the material~\cite{jung2021effect}. However, despite much interest in VHSs, there are several challenges to realizing an electric-field-controlled system. First, the system should be stable in air for practical application. Second, VHSs should be placed near $E_{\rm F}$. Third, the material should be two-dimensional (2D) to ensure that the DOS diverges (Fig. 1a)~\cite{van1953occurrence}. Finally, the external electric field should be strongly screened by charge carriers in a metallic system for a controllable region to be confined near the surface~\cite{thomas1927calculation,nakano2012collective,ahn2003electric,nelson2022interfacial}. Due to these various challenges, only a few controllable and practical devices have been realized and reported.

To overcome the above issues and fully exploit the functionality of VHSs, we utilize an atomically ultra-thin SrRuO$_3$ (SRO) film with a 2D VHS located near $E_{\rm F}$~\cite{shen2007evolution,kim2022heteroepitaxial}. Ultra-thin and superlattice film systems have been extensively studied due to their strong tunability~\cite{nelson2022interfacial,kim2022heteroepitaxial,mori2019controlling,yukawa2021resonant,king2014atomic,monkman2012quantum}. In this study, we experimentally control a VHS with Lifshitz transition (Fig. 1b) and tunable transport properties by dosing the surface with potassium (K) (Fig. 1c) and applying the ionic liquid gating method (Fig. 1d). Our platform can be a new strategy to fully exploit the functionality of VHS systems, as it overcomes the difficulty in tuning physical properties of metallic systems.

%%%%%%%%%%%%%%%%%%
\begin{figure*}[htbp]
	\includegraphics[width=0.8\textwidth]{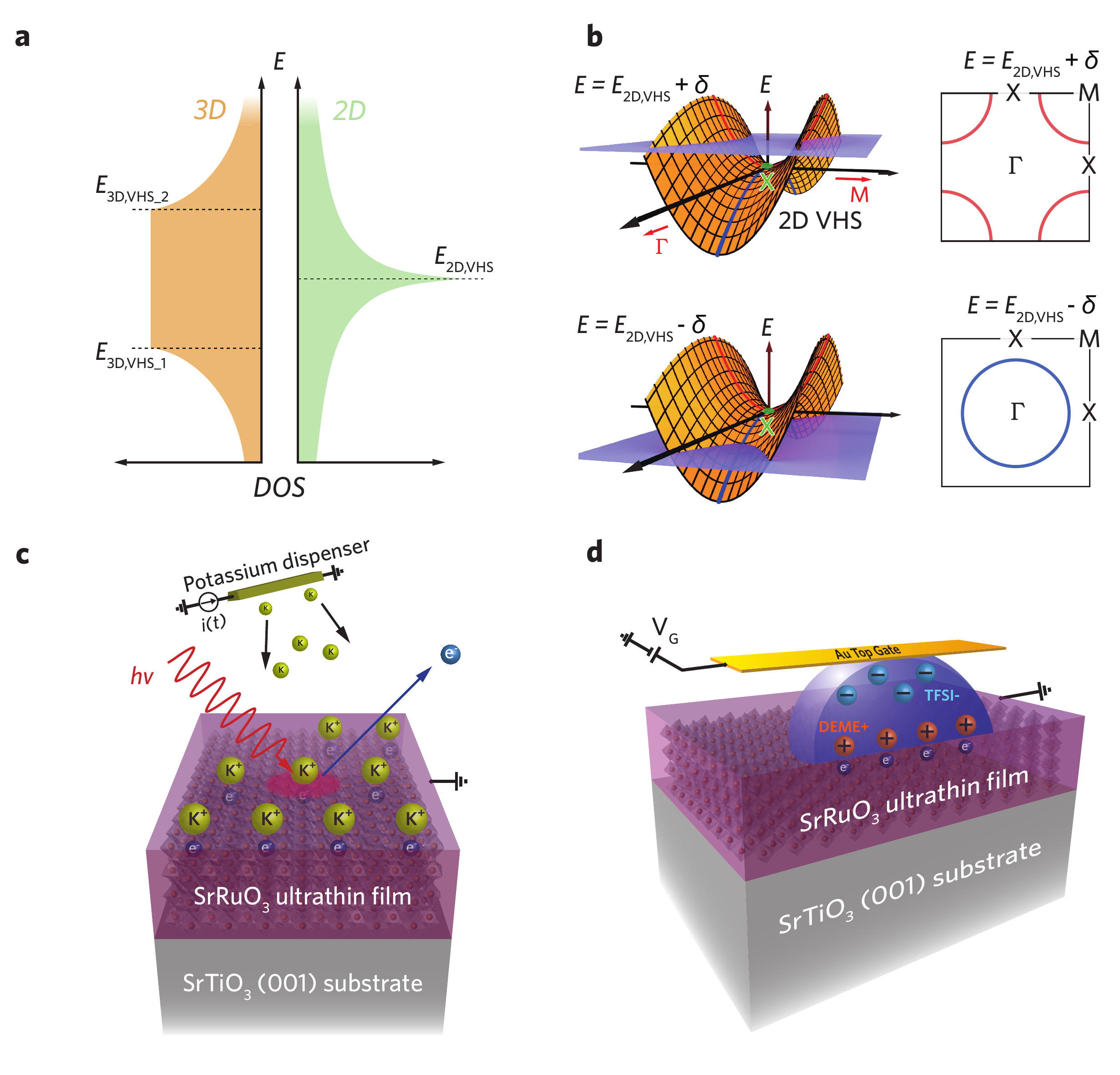}
	%\vspace{-1cm}
	\caption{{\bf Tunable two-dimensional (2D) van Hove singularities (VHSs) with applied electric fields.} {\bf a} Density of states (DOS) near the VHSs in three-dimensional (3D) and 2D systems. DOS only diverges in the 2D VHS system at $E_{\rm 2D,\!VHS}$. $E_{\rm 3D,\!VHS}$ ($E_{\rm 2D,\!VHS}$) represents an energy 3D (2D) VHS is placed at. {\bf b} Schematic diagrams of the 2D VHS and corresponding constant energy maps below and above the VHS, $E=E_{\rm 2D,\!VHS} \pm \delta$, where $\delta$ is the small variation of the chemical potential.  {\bf c} Schematic illustration of the alkali metal dosing method used to apply an external electric field {\it in-situ}. Potassium $\rm (K)$ atoms are deposited on the surface which induces an electric field and electrons near the surface. {\bf d} Schematic illustration of the ionic liquid gating method used to apply an external electric field {\it ex-situ}. The applied electric field can be controlled with ionic liquid.}
	\label{fig:1}
\end{figure*}
%%%%%%%%%%%%%%%%%%%%%%%%%%%%%%%%%%%%%%%%%%%%%%%%%%%%%%%%%%%

\thispagestyle{fancy}%

\section*{Results}

A 4 unit-cell (uc) SRO ultra-thin film was grown using a pulsed laser deposition method (See Methods for details). Subsequently, the sample was transferred \emph{in-situ} and analyzed by angle-resolved photoemission spectroscopy (ARPES). Schematics of the Fermi surfaces of the 2D SRO (Figs. 2a-c) indicate that the VHS of the $\gamma$ band at the $X$ point should be located near $E_{\rm F}$, based on previously reported band structures~\cite{sohn2021observation,sohn2021sign}. When the VHS is placed above (below) $E_{\rm F}$, the $\gamma$ band is expected to have an electron (hole)-like Fermi surface (Fig. 1b). We analyzed the Fermi surfaces of a 4~uc SRO thin film (Figs. 2d and 2h) to determine which one corresponds to the actual shape of the Fermi surface. Two bands cross $E_{\rm F}$ along the $\Gamma$-$X$ direction in the first Brillouin zone ($k < 0.8~\rm \AA^{-1}$), where the inner band is the $\beta$ band and the outer band near the zone boundary ($k = 0.8~\rm \AA^{-1}$) is the $\gamma$ band. Since the Fermi momentum ($k_{\rm F}$) of the $\gamma$ band is located on the $\Gamma$-$X$ cut in the first Brillouin zone, the pristine 4~uc SRO has an electron-like $\gamma$ band~\cite{sohn2021observation} (Fig. 2a).

To track the VHS as a function of the chemical potential, we dosed the surface of the 4~uc SRO thin film with K. Alkali metal dosing (AMD) is an effective way to change the chemical potential, where the difference in work function between alkali metal and dosed system induces a strong electric field on the surface~\cite{jung2021effect}. K was deposited several times, and ARPES measurements were conducted between the depositions, starting from the pristine state (see Methods for the detailed K dosing sequence and parameters). Figs. 2d-g show Fermi surface maps with various dosing levels. The size of the $\beta$ band pocket increases upon K dosing, which demonstrates that K dosing effectively adds electrons to the system. The VHS of the $\gamma$ band near the $X$ point passed through $E_{\rm F}$; indicating a Lifshitz transition of the $\gamma$ band (Figs. 2d-g). Therefore, the $\gamma$ band of the pristine sample is an electron-like band centered around the $\Gamma$ point (Fig. 2a), but becomes a hole-like band centered around the $M$ point with K dosing (Fig. 2c).

The electron doping-dependent tendency is captured in the dosing-dependent high symmetry cuts along the $\Gamma$-$X$-$\Gamma$ (Figs. 2h-k). In the pristine sample, the $\gamma$ band crosses $E_{\rm F}$ along the $\Gamma$-$X$ high symmetry line, which implies that the $\gamma$ band is an electron-like band, as shown in Fig. 2a~\cite{note}. After K dosing, the top of the $\gamma$ band at the $X$ point is clearly below $E_{\rm F}$ (Fig. 2k), indicating that the $\gamma$ band turns into a hole-like band, as illustrated in Fig. 2c. In addition, the extracted momentum distribution curves (MDCs) in Fig. 2l show that the $\beta$ band moves towards the zone boundary ($X$ point). We noticed that the surface of the “Dosing 3” SRO is almost covered with a monolayer of K, since the position of $\beta$ band is saturated in the further dosed sample (See Supplementary Materials (SM) for dosing-dependent $\beta$ band positions). Note that the $\gamma$ band is not very clear along $k_y$~=~0 (Fig. 2h) due to the matrix element effect~\cite{sohn2021observation,hahn2021observation}. However, it can be clearly seen along $k_x$~=~0 (see SM for details).

%%%%%%%%%%%%%%%%%%
\begin{figure*}[htbp]
	\includegraphics[width=1\textwidth]{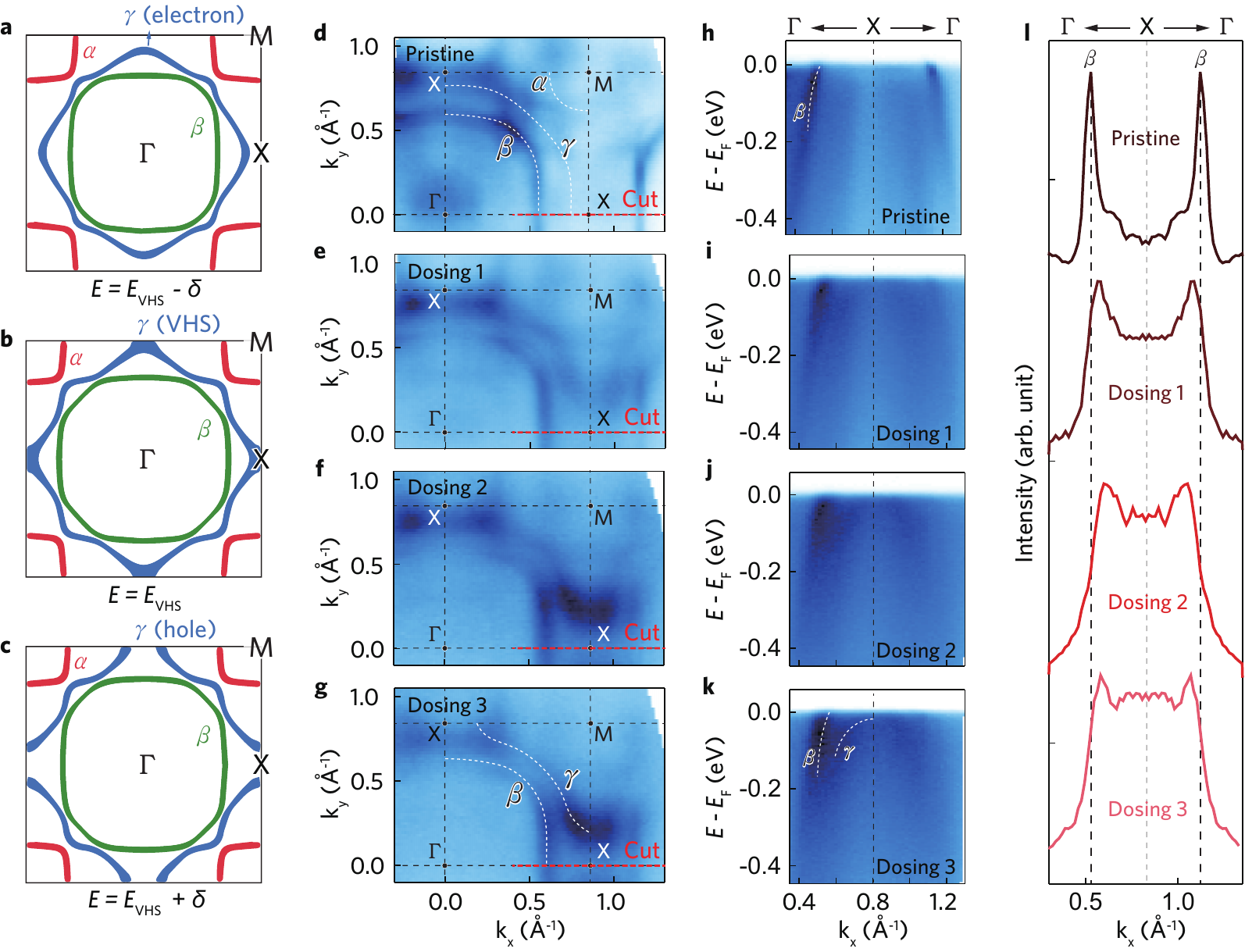}
	%\vspace{-1cm}
	\caption{{\bf Electron band control of the 2D VHS with K deposition.} {\bf a-c} Fermi surface schematics of 4~uc SRO thin films with the small variation ($\delta$) of the chemical potential. The electron- and hole-like $\gamma$ bands are described in {\bf a} and {\bf c}, respectively. {\bf d-g} Fermi surfaces of a pristine/K-dosed SRO thin film measured with angle-resolved photoemission spectroscopy (ARPES). "Dosing {\it n}" indicates that K atoms are deposited with {\it n} sets on the surface of the SRO thin film. The black dashed lines denote high symmetry lines of the tetragonal Brillouin zone of SRO. {\bf h-k} High symmetry cuts along the $\Gamma$-$X$-$\Gamma$ direction (cuts in {\bf d-g}). Black dashed lines denote the zone boundary at the $X$ point. {\bf l} Momentum distribution curves (MDCs) at the Fermi level ($E_{\rm F}$) along the $\Gamma$-$X$-$\Gamma$ high symmetry cut. Black dashed lines denote the $\beta$ band position of the pristine sample. The MDC is symmetrized with respect to the $X$ point. The grey dashed line denotes the $X$ point. The intensity offsets of the curves are set arbitrarily for better visualization. }
	\label{fig:2}
\end{figure*}
%%%%%%%%%%%%%%%%%%%%%%%%%%%%%%%%%%%%%%%%%%%%%%%%%%%%%%%%%%%

Transport properties are expected to change with variation of the chemical potential as the VHS is placed near $E_{\rm F}$. A sign change can be expected in the majority charge carrier type, from electron- to hole-type. This is due to the electron-like $\gamma$ band turning into a hole-like band with increasing the chemical potential. However, AMD is not suitable for transport measurements, which are conducted in ambient conditions whereas AMD requires high vacuum conditions~\cite{aruga1989alkali}. A similar method used to change the chemical potential is ionic-liquid gating which is widely adopted for transport studies of ultra-thin systems~\cite{kim2021capping,nakano2012collective}. Therefore, we utilized an ionic-liquid gating method (Fig. 1c) and performed transport measurements with variation of gate voltage and temperature~\cite{sohn2021sign,bisri2017endeavor,lin2021electric} for systemic control of the chemical potential in ambient conditions.

%%%%%%%%%%%%%%%%%%
\begin{figure*}[htbp]
	\includegraphics[width=1\textwidth]{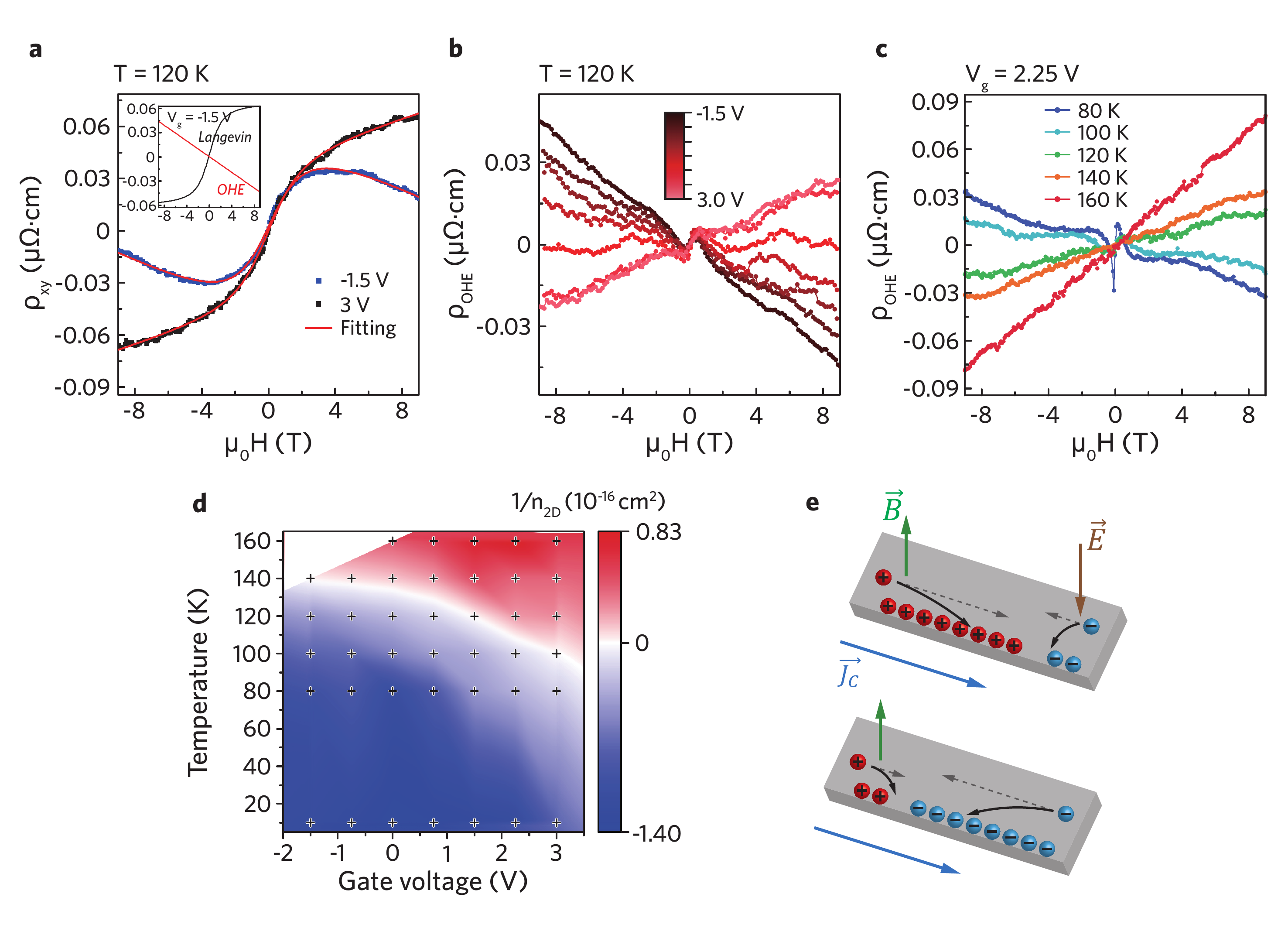}
	%\vspace{-1cm}
	\caption{{\bf Transport control of the 2D VHS with ionic-liquid gating.} {\bf a} Hall resistivity and fitted data with the gate voltage from -1.5 to 3~V at the temperature of 120~K. The fitted data represent the superposition of the Langevin function and ordinary Hall effect (OHE) contribution. Fitted Langevin function and ordinary Hall resistivity ($\rho_{\rm OHE}$) are separately plotted in the inset. {\bf b} $\rho_{\rm OHE}$ is plotted as a function of gate voltage from -1.5 to 3~V at 120~K. {\bf c} $\rho_{\rm OHE}$ is plotted as a function of temperatures under a gate voltage of 2.25~V. {\bf d} Contour plot of the inverse of charge carrier density, ($1/n_{\rm 2D}$), with variation of temperature and gate voltage; black crosses denote the experimentally measured points. $n_{\rm 2D}$ diverges in the white region. {\bf e} Schematic illustration of tunable charge carriers during transport measurements. E, B, J$_C$ represent the electric field applied by ionic liquid gating, magnetic field and current density for Hall measurements, respectively.}
	\label{fig:3}
\end{figure*}
%%%%%%%%%%%%%%%%%%%%%%%%%%%%%%%%%%%%%%%%%%%%%%%%%%%%%%%%%%%

Figure 3a shows the Hall resistivity of the 4~uc SRO thin film, measured at 120~K with gate voltages between -1.5~V and 3~V. The Hall resistivity in SRO thin films is well described by three terms, i.e., $\rho_{xy} = \rho_{\rm OHE} + \rho_{\rm AHE} + \rho_{\rm hump}$, which are the Hall resistivities of the ordinary Hall effect, anomalous Hall effect, and a hump-like feature, respectively~\cite{sohn2021stable,kim2021capping,sohn2020hump}. When the temperature is near the Curie temperature, $T_{\rm C}\sim113~{\rm K}$ (see SM for a definition of the Curie temperature), thermal fluctuation induces spin chirality fluctuation, and curved-shape Hall hysteresis is observable in the low magnetic field region~\cite{wang2019spin,majcher2014magnetic}. To obtain $\rho_{\rm OHE}$, we fitted the results with the Langevin and linear functions to extract the ordinary Hall contribution from the non-linear experimental data~\cite{wang2019spin}. Thus, the $\rho_{\rm OHE}$ component can be extracted from $\rho_{\rm xy}$ after fitting with the Langevin function, as shown in Fig. 3a (inset).

We plotted $\rho_{\rm OHE}$ over a range of gate voltages and temperatures. $\rho_{\rm OHE}$ exhibits voltage dependency, which means that the slope of the carrier density changes with gate voltages, as shown in Fig. 3b. When the gate voltage is -1.5~V, the slope is negative, but becomes positive as the gate voltage increases. A similar tendency has also been observed with temperature variations. Figure 3c shows the temperature-dependent $\rho_{\rm OHE}$ when a fixed 2.25~V gate voltage is applied. As the temperature increases, the slope of $\rho_{\rm OHE}$ changes from negative to positive. 

To determine the relation between the slope in $\rho_{\rm OHE}$ and the VHS, we extracted the inverse of carrier density, i.e., $1/n_{\rm 2D}$, defined as $1/n_{\rm 2D}$ = $R_He/t$, where $R_H$, $e$, and $t$ are the slope of Hall resistivity (or, Hall coefficient), an elementary charge, and thickness, respectively~\cite{kim2021capping}. $n_{\rm 2D}$ is proportional to the DOS near  $E_{\rm F}$, i.e., $n_{\rm 2D}\propto \int_{E_{\rm F}-\Delta}^{E_{\rm F}+\Delta} \! g(E) \, \mathrm{d}E$, where $g(E)$ is DOS. According to these equations, $n_{\rm 2D}$ diverges as the slope of $\rho_{\rm OHE}$ goes to zero when the 2D VHS becomes close to $E_{\rm F}$. When $E_{\rm F}$ crosses the 2D VHS, the slope of $\rho_{\rm OHE}$ changes its sign, resulting in a sign change of $1/n_{\rm 2D}$. In other words, the charge carrier diverges and changes its sign with small variation of the chemical potential. Figure 3d shows a contour plot of $1/n_{\rm 2D}$ by temperature and gate voltage. Near the VHS, the DOS diverges when $1/n_{\rm 2D}$ goes to zero, as the white-colored region indicates, and a sign change of the majority charge carrier follows (Fig. 3e).

\section*{Discussion}

%%%%%%%%%%%%%%%%%%
\begin{figure*}[htbp]
	\includegraphics[width=1\textwidth]{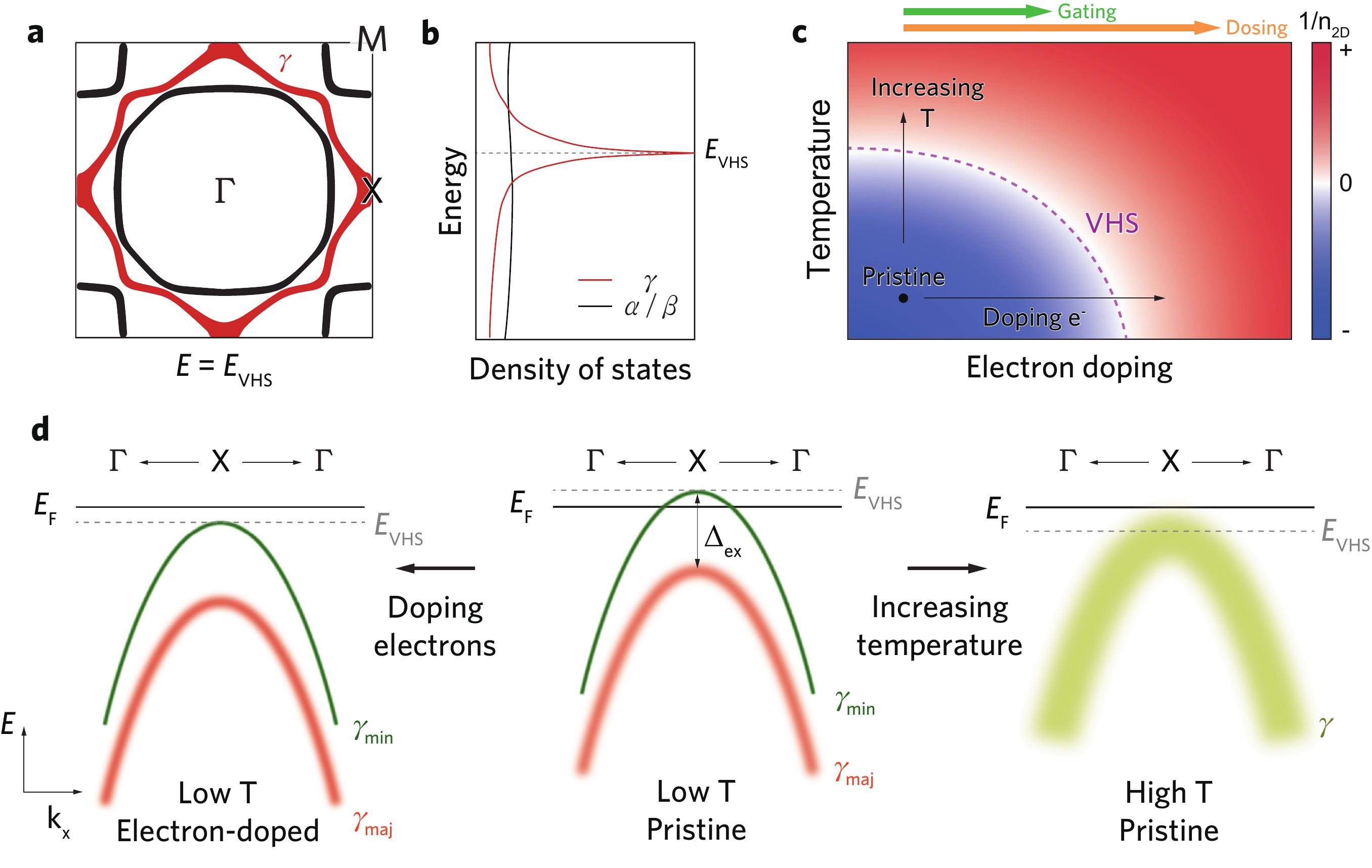}
	%\vspace{-1cm}
	\caption{{\bf The 2D VHS in the effective one band model and ferromagnetism in SRO.} {\bf a} Schematic of the SRO Fermi surface when $E_{\rm F}$ crosses the 2D VHS. {\bf b} Schematic of the DOS of the $\gamma$ (red) and $\alpha/\beta$ (black) bands. {\bf c} Schematic contour plot of $1/n_{\rm 2D}$ as a function of temperature and electron doping. $n_{\rm 2D}$ diverges when $E_{\rm F}$ crosses the VHS (denoted by the purple dotted line). {\bf d} Band diagrams of a ferromagnetic VHS model along the $\Gamma$-$X$-$\Gamma$ high-symmetry line. When $E_{\rm VHS}$ is placed higher (lower) than $E_{\rm F}$, the $\gamma$ band is electron-like (hole-like) at the Fermi surface. The green and red bands represent $\gamma$ minority and majority bands, respectively. $\Delta_{\rm ex}$ denotes the exchange splitting energy between the $\gamma$ minority and majority bands.}
	\label{fig:4}
\end{figure*}
%%%%%%%%%%%%%%%%%%%%%%%%%%%%%%%%%%%%%%%%%%%%%%%%%%%%%%%%%%%

We have experimentally shown that the 2D VHS of the SRO film can be placed at $E_{\rm F}$ by electron doping, and that the Hall coefficient changes significantly as a function of temperature and electron doping. In this section, we will focus on how the electronic structures and VHS determine the Hall coefficient of the SRO film.

First, we discuss how the VHS can determine Hall coefficients in a multi-band system. The Fermi surface of 4~uc SRO thin films  shows multiple pockets at $E_{\rm F}$ as shown in Fig. 2d, and each band crossing $E_{\rm F}$ is expected to contribute to the Hall coefficient. The Hall coefficient of a multi-band system can be calculated from a weighted sum of their carrier densities and mobilities, as follows:~\cite{allgaier1967high,ashcroft1976solid},
\begin{equation*}
R_{H, total} = \frac{\sum_j \! \frac{n_je_j\mu_j}{1+\mu_j^2H^2}}{({\sum_j \! \frac{n_je_j\mu_j}{1+\mu_j^2H^2}})^2+({\sum_j \! \frac{n_je_j\mu_j^2H}{1+\mu_j^2H^2}})^2H^2},
\end{equation*}
where $j$ is the band index, $n_j$, $\mu_j$, $e_j$, and $H$ denote the carrier density, carrier mobility, charge of the corresponding bands, and the external magnetic field, respectively. Thus, determining the total Hall coefficient and a majority charge carrier is not a trivial issue in multi-band systems. However, when the multi-band system has a 2D VHS, the DOS diverges in proximity to the VHS (Fig. 1a). Hence, the carrier density of the VHS band is dominant over the other bands (Figs. 4a and 4b) for $n_{\rm VHS} \gg n_j$ case, the Hall resistance can be approximated to an effective one band model as $R_{\rm H,total} \approx \frac{1}{n_{\rm VHS}e_{\rm VHS}}$. This suggests that the Hall coefficient is dominated by the VHS band when the system is in proximity to the VHS, resulting in a zero-slope Hall coefficient regardless of the other bands. Due to the VHS of the $\gamma$ band crossing $E_{\rm F}$ upon electron doping, the $\gamma$ band of SRO exclusively determines the Hall coefficients due to the divergent DOS, as shown in Fig. 4b.

Next, we compare the ionic-liquid gating and AMD methods. The two methods share the same mechanism, i.e., introduce charge carriers near the surface of the sample. However, a key difference is the amount of induced charges. The electric field produced by the alkali metal on the SrO surface is of the order $1~\rm {V/\AA}$~\cite{kyung2021electric}, while that produced by liquid gating is of the order $0.1~\rm {V/\AA}$~\cite{bisri2017endeavor,saito2016gate,hwang2012emergent}. Thus, the AMD method should induce an order of magnitude more charge carriers into the system compared to the ionic-liquid gating method, which would result in a significantly larger change in chemical potential. For $1/n_{\rm 2D}$, a change in the majority of carrier types was not seen at low temperatures, while the ARPES results showed a clear Lifshitz transition in the $\gamma$ band. Based on the above, a schematic of the extended phase diagram of $1/n_{\rm 2D}$ is provided in Fig. 4c.

Finally, we discuss the temperature dependence of $1/n_{\rm 2D}$. The Hall coefficients shown in Fig. 3 exhibit crossover of the Hall coefficient signs with changes of temperature and chemical potential. The temperature-dependent tendency might be attributed to the temperature-dependent exchange splitting of the $\gamma$ majority and minority bands. Previous dynamic mean-field theory and ARPES studies of SRO systems~\cite{hahn2021observation,kim2015nature} revealed clear signatures for the exchange splitting, and coherence of the minority band. Since minority bands are only present in coherent states~\cite{hahn2021observation,kim2015nature}, we speculate that the $\gamma$ band observed in our ARPES data is a $\gamma$ minority band. Considering that exchange splitting monotonically decreases upon heating~\cite{hahn2021observation,jeong2013temperature}, the $\gamma$ minority band should move toward higher binding energy. Thus, the VHS, which was originally located above $E_{\rm F}$ at low temperatures reaches  $E_{\rm F}$ and moves  below $E_{\rm F}$ at high temperatures. Figure 4d provides schematics of the band structures of the SRO along the $\Gamma$-$X$-$\Gamma$ direction. The VHS of the $\gamma$ band, which was placed above $E_{\rm F}$, moves below  $E_{\rm F}$ with increasing temperature and doping electrons, which accounts for the temperature- and electron doping-dependent crossovers of $1/n_{\rm 2D}$ shown in Fig. 4c.

In summary, we have experimentally shown that ionic-liquid gating and AMD are effective ways to control the 2D VHS in metallic ultrathin films, which can host flat bands. In the ultra-thin limit, a strong electric field on the surface can change the physical properties of metallic systems. The proposed platform provides a new route to realizing flat bands near $E_{\rm F}$, allowing us to engineer electronic correlations without changing the sample. We believe that our findings broaden the potential applications of flat band systems, and should facilitate further studies on electronic correlations engineering by external stimuli in ultra-thin film systems.

\section*{Methods}
{\bf SRO thin film fabrication.} Epitaxial SrRuO$_3$ (SRO) thin films were grown on atomically flat TiO$_2$-terminated SrTiO$_3$ (001) (STO) substrates by pulsed laser deposition (KrF; 248~nm wavelength). Before film deposition, STO substrates were sonicated in deionized water for 30~minutes, and pre-annealed {\it in-situ} at 1070~$^{\circ}{\rm C}$ under oxygen partial pressure of $5\times 10^{-6}$~Torr. SRO films were deposited at 670~$^{\circ}{\rm C}$ under oxygen partial pressure of 100~mTorr. The energy fluence and repetition rate of the excimer laser were set to 2~${\rm J/cm^2}$ and 2~Hz, respectively. Reflection high-energy electron diffraction (RHEED) was used to monitor the growth process {\it In-situ}.

\hfill

{\bf ARPES measurements.} For the photoemission experiments, the sample was transferred {\it in-situ} to the preparation chamber to obtain a clean surface by post-annealing at 550~$^{\circ}{\rm C}$ for 10~minutes. Subsequently, the sample was transferred {\it in-situ} to the angle-resolved photoemission spectroscopy (ARPES) chamber. Measurement was conducted with He-I$\alpha$ ($h\lambda= 21.2~eV$) at 10~K using an ARPES system equipped with a DA30 analyzer (Scienta Omicron) and a discharge lamp (Fermi instruments). He-I$\alpha$ ($h\lambda= 21.2~eV$) light was used. The base pressure of the ARPES chamber was kept better than $3\times 10^{-11}$~Torr. After obtaining ARPES data from the pristine SRO, we deposited potassium (K) on the surface of the SRO at 10~K using an alkali metal dispenser (SAES getters). An electric current of 5.4~A was applied to the K dispenser for 2~minutes om each dosing step.

\hfill

{\bf Fabrication of ionic liquid gating device.} For the transport measurements, 60~nm Au electrodes were deposited on SRO films using an e-beam evaporator. Electric contacts on the samples were made by an ultrasonic wire bonder with Al wires. A low vapor pressure epoxy (Torr Seal; Varian) was used to protect electrical contacts and surround the ionic liquid on the device. An Au plate was used as a top gate electrode.

\hfill

{\bf Transport measurements.} Hall effect measurements were conducted using a physical property measurement system (PPMS; Quantum Design). Diethylmethyl(2-methoxyethyl)ammonium bis(trifluoromethylsulfonyl)imide (DEME-TFSI) was used for ionic liquid gating. Before the measurements, the ionic liquid (DEME-TFSI) was baked at 110~$^{\circ}{\rm C}$ for 24~hours in an Ar-filled globe box to remove water remaining in the ionic liquid. The ionic liquid was dropped on the device immediately before loading the sample into the PPMS to minimize water contamination. Gate voltage was adjusted at 220~K and kept for 30~minutes to stabilize the ionic liquid.

\acknowledgments
The authors wish to thank B. Seok for fruitful discussions. This work was supported by the Institute for Basic Science in Korea (Grant No. IBS-R009-G2) and the National Research Foundation of Korea (NRF) grant funded by the Korea government (MSIT). (No. 2022R1A3B1077234). 

\section*{data availability}
The data that support the findings of this study are available from the corresponding author upon reasonable request.

\ifarXiv


\section*{Supplementary material}

**Supplementary Materials for**

# Electric control of two-dimensional Van Hove singularity in oxide ultrathin films

**Supplementary Note 1: Characterization of SrRuO₃ (SRO) ultrathin films**

Fig. S1 shows the information on the 4 uc SRO film studied in this work. The atomic force microscopy topographic image of the SRO shown in Fig. S1a shows that the SRO has an atomically flat surface, which is suitable for angle-resolved photoemission spectroscopy (ARPES) and transport measurements. The reflection high energy electron diffraction image in Fig. S1b shows that the film was grown in a step-flow growth mode after layer-by-layer growth of 2 uc. Figs. S1c and d show the resistivity and Hall effect measurement results of the pristine SRO, respectively. The inset of Fig. S1c is the temperature derivative of the resistivity, showing a Curie temperature of 113 K. The Hall resistivity reproduces the anomalous Hall effect and hump-like features, which were reported in the previous studies[1,2].

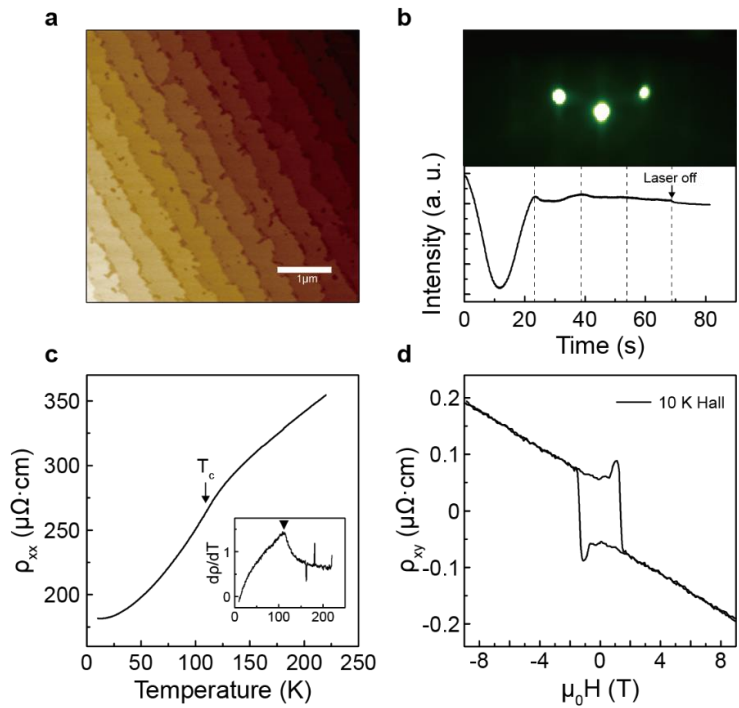

**Fig. S1. Characterization of 4 unit cell (uc) SrRuO₃ (SRO) film. a.** Atomic force microscopy image of 4 uc SRO film grown on SrTiO (001) substrate. **b.** *In-situ* monitored reflection high energy electron diffraction pattern and corresponding intensity of the center spot during 4 uc SRO growth. **b.** Temperature dependent resistivity of 4 uc SRO film. Inset, temperature derivative of the resistivity. Curie temperature of the SRO film ($T_c \sim 113$ K) is determined from the peak position of the resistivity derivative. **d.** The $\rho_{xy}$ of 4uc SRO film at 10 K.

**Supplementary Note 2: Estimation of potassium (K) coverage on SRO**

The K coverage on the SRO surface is determined by dosing-dependent β band positions. As K is deposited on the surface of the SRO, electrons are introduced to the SRO, resulting in the shift of the bands toward higher binding energy. Therefore, the peak-to-peak distance of the β bands of the first and second Brillouin zone ($\Delta k_\beta$) at the Fermi level can be a measure of electron occupations (see Fig. S2a). On the other hand, it has been reported that the K coverage of more than 1 monolayer does not introduce additional electrons into the system[3,4]. As can be seen in Fig. S2b, the $\Delta k_\beta$ gradually decreases with K deposition, but becomes saturated at the dosed sets of 3, indicating that the surface of the sample is covered with almost 1 monolayer K coverage at the dosed sets of 3.

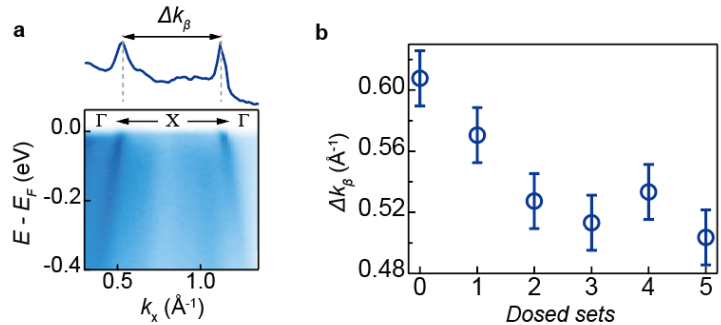

**Fig. S2. Dosing dependent β band positions. a.** High symmetry cut of the pristine SRO along the Γ-X-Γ direction and extracted momentum distribution curve at $E_F$. $\Delta k_\beta$ is defined as the peak-to-peak distance of the β bands of the first and second Brillouin zone. **b.** $\Delta k_\beta$ as a function of conducted sets of the K dosing on the SRO surface. Error bars denote the momentum resolution of the ARPES measurements.


**Supplementary Note 3: γ band topology of the pristine SRO**

The γ band is not well resolved in the Γ-X cut of the pristine SRO along the x-direction (perpendicular to the analyzer slit) due to the matrix element effects. To figure out whether the γ band of the pristine SRO is an electron- or hole-like band, we plotted another Γ-X cut along the y-direction (parrel to the analyzer slit) in Fig. S3. As can be seen, the γ band crosses the Fermi level along the Γ-X direction, indicating the band is an electron-like band.

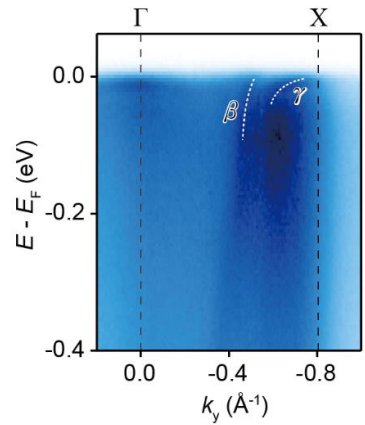

**Fig. S3.** High symmetry cut of the pristine SRO along the axis perpendicular to the analyzer slit.


**Supplementary Note 4: Results of ionic liquid gating on 4 uc SRO films**

Hall resistance of 4 uc SRO films as a function of the magnetic field, temperature, and gate voltage is shown in Fig. S4. The sign of anomalous Hall effect changes as a function of gate voltage, as previously reported[2]. In addition, the hump-like feature in the Hall resistance can be seen. In the Hall resistance data measured at temperatures above 80 K, the anomalous Hall contribution does not saturate up to a magnetic field of 9 T.

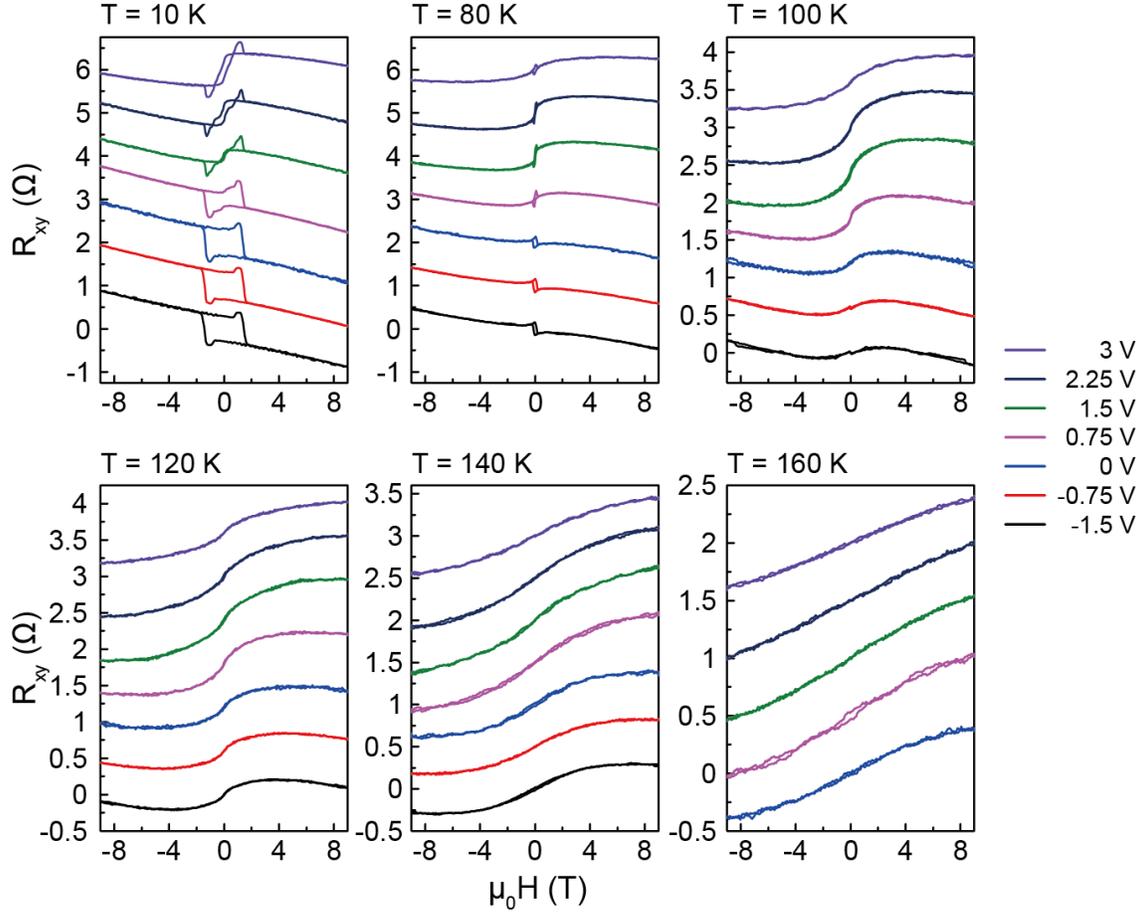

**Fig. S4. Hall resistance of 4 uc SRO film as a function of magnetic field, temperature and gate voltage.**

**Supplementary Note 5: Extracting anomalous Hall contribution from Hall resistance**

The magnetization curve of anomalous Hall contribution follows the Langevin function[5]. To extract the ordinary Hall contribution which is expected to follow a linear function, the fitting function was set as follows,

$$y = y_0 + p_1 \left[ 1/\tanh\frac{x-x_0}{p_2} - 1 \Big/ \frac{x-x_0}{p_2} \right] + p_3 x$$

where $x_0, y_0, p_1, p_2, p_3$ are fitting parameters.

The terms in the brackets are from the Langevin function, and the last term, $p_3 x$, is the ordinary Hall contribution. The ordinary Hall resistivity results in Fig. 3 of the manuscript are obtained by subtracting the Langevin function from each measurement with the fitted parameters $x_0, p_1, p_2$.

**Supplementary Note 6: Carrier density in transport experiments**

The carrier density is estimated from the slope of the ordinary Hall contribution. Shown in Fig. S5 are carrier density $n_{2D}$ as a function of temperature and gate voltage. The positive (negative) sign of the carrier density corresponds to electrons (holes).

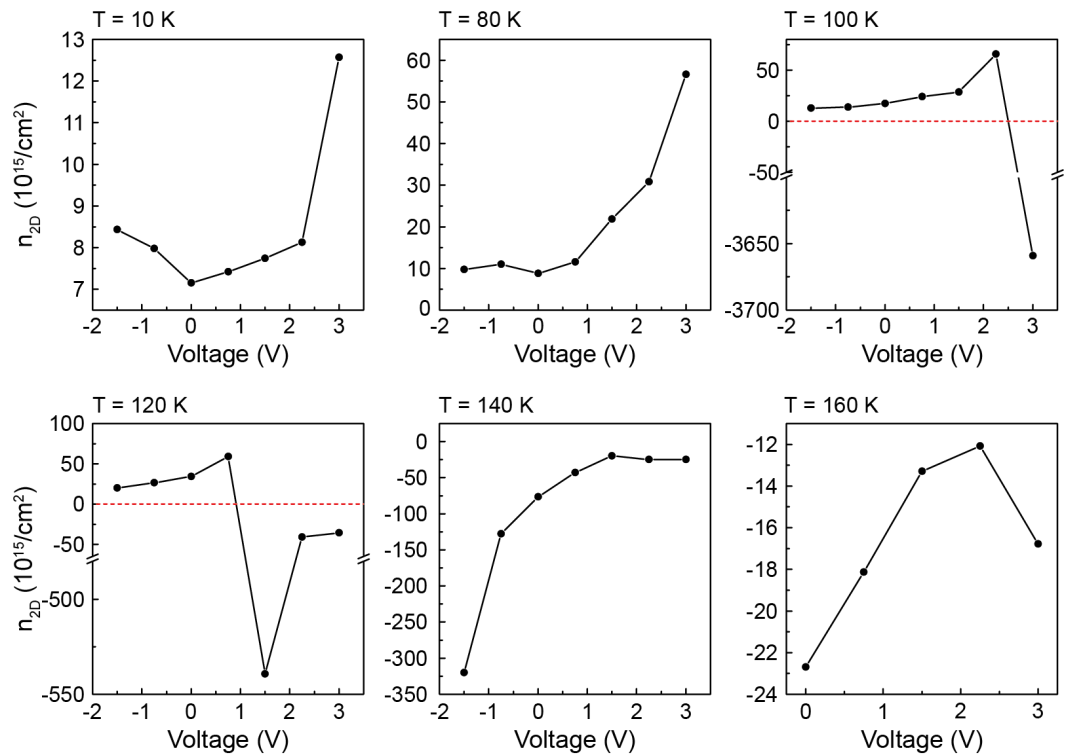

**Fig. S5. Carrier density obtained from Hall effect measurements.** Carrier density as a function of the applied gate voltage and temperature.


\fi

\end{document}